# TCP PERFORMANCE FOR KURD MESSENGER APPLICATION USING BIO-COMPUTING

**Bnar Faisal Daham**
Software Engineering Dept/ College of Engineering - Salahaddin University

**Ayad Ghany Ismaeel**
Information System Engineering Dept
Erbil Technical Eng. College

**Suha A. Abdual-Rahman**
Software Engineering Dept
College of Engineering/ Salahaddin University

**Abstract**

This work was conducted to design, implement, and evaluate a new model of measuring Transmission Control Protocol (TCP) performance of real time network. The proposed model Biological Kurd Messenger (BIOKM) has two main goals: First is to run the model efficiently, second is to obtain high TCP performance via real time network using bio-computing technique, especially molecular calculation because it provides wisdom results and it can exploit all facilities of phylogentic analysis.

To measure TCP performance two protocols were selected Internet Relay Chat Daemon (IRCD) and File Transfer Protocol (FTP), the BIOKM model consists of two applications Kurd Messenger Server Side (KMSS) and Kurd Messenger Client Side (KMCS) written in Java programming language by implementing algorithms of BIOKM Server and Client application. The paper also includes the implementation of hybridized model algorithm based on Neighbor-Joining (NJ) method to measure TCP performance, then implementing algorithm of Little's law (steady state) for single server queue as a comparison with bio-computing algorithm. The results obtained by using bio-computing and little's law techniques show very good performance and the two techniques result are very close to each other this is because of local implementation. The main tools which have been used in this work can be divided into software and hardware tools.

**Keywords:** Biological Kurd Messenger (BIOKM), Kurd Messenger Phylogenetic tree, Hybridized Model, Little's Law, TCP Performance.

## Introduction

There are multiple types of Internet Protocol IP Network which is based on Transmission Control Protocol/Internet Protocol TCP/IP Model such as:

**i.** The internet which is a global information system consisting of millions of computer networks around the world. It is also a common and ordinary infrastructure for all kinds of human activities, especially research and education applications benefit from high-speed long-distance connection

**ii.** The Technology can be applied (purely internal users) as Intranet, which is used Firewall to make accessible only to members of enterprise,

**iii.** The Extranet are extended version of Intranets, which allowed the authorized to access only.

The TCP and the User Datagram Protocol UDP are both IP transport-layer protocols. UDP is a lightweight protocol that allows applications to make direct use of the unreliable datagram service provided by the underlying IP service. UDP is commonly used to support applications that use simple query/response transactions, or applications that support real-time communications. TCP provides a reliable point-to-point communication channel that clients and servers can use to communicate with one another. For each IP network the measurement of performance is very important, so the TCP can used as performance measuring because it guarantees send data from one end of the connection actually gets to the other end and in the same order it was send. Otherwise, an error is reported. The Hypertext Transfer Protocol HTTP, FTP, IRCD, and Telnet are all examples of applications that require a reliable communication channel TCP [1].

---

\* Cited from M.Sc. thesis.

There are several methods to measure the performance of the TCP among these methods genetic algorithms, neural network, data mining, and bio-computing. For this paper, data mining, genetic, neural network were not used for their weakness in this case.

The definitions of bio-computing will start with two basic principles: "information" and "control" to build a definition based on them by means of interaction with five related basic science disciplines **[2]**:

i. Physics: Includes physical components (physical material) and physical factors (such as temperature).
ii. Biology and Philosophy: Application of bio-computing application.
iii. Mathematics: Mathematical model (linear equation, and equivalence…).
iv. Engineering: Includes genetic engineering, software engineering (the establishment of sound engineering in order to obtain a reliable software which work efficiently on real machine with low cost), and encapsulation (the addition of control information by a protocol entity to data obtained from protocol user).

The fields of information and control can be interrelated, as shown in figure (1). The relationships between these disciplines information and control as bases of bio-computing, give five entities as a result of the synthesis, of the five disciplines and the basic elements of bio-computing.

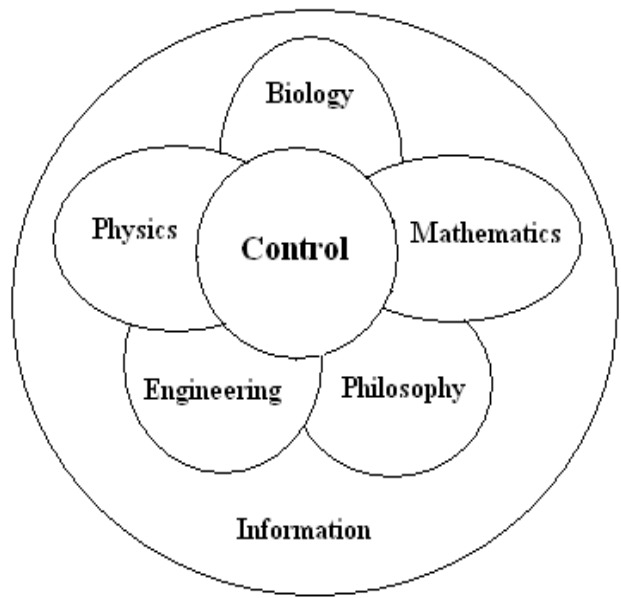

**Figure 1: BIO-Computing Science Interconnection**

The synthesized entities which can constitute bio-computing can be stated bellow:
i. Bioinformatics: Is the field of science in which biology, computer science, and information technology merge to form a single discipline bio-medical informatics.
ii. Molecular computing: Advanced technology for (Computer science application, and mathematics).
iii. Information biotechnology: It is the synthesis between information technology and biotechnology to create an advanced technology within a specific field of application as shown in figure (2).
iv. Bio-computing philosophy.

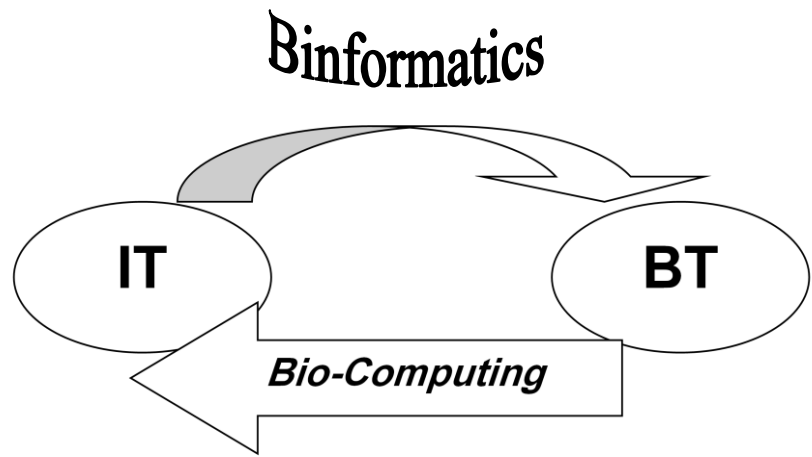

**Figure 2: Information Biotechnology**

The problem of this work is to measure the network performance using bio-computing technique. The TCP has been selected for this purpose because TCP provides a communication channel between processes on each host system. The channel is reliable, full-duplex, and streaming. To achieve this functionality, the TCP drivers break up the session data stream into discrete segments, and attach a TCP header to each segment. An IP header is attached to this TCP packet, and the composite packet is then passed to the network for delivery. This TCP header has numerous fields that are used to support the intended TCP functionality.

The best technique of solving the problem of TCP performance measurement is by using Bio-computing, especially molecular calculation which is used to evaluate and study the TCP performance, because it provides wisdom results and it can exploit all facilities of phylogenetic analysis.

There are many phylogenetic analysis methods and their strategies as shown in table (1). Neighbor Joining NJ method was selected for this purpose because the NJ constructs the tree by sequentially finding pairs of neighbors, which are the pairs of Operational Taxonomy Unit OTUs connected by a single interior node.

**Table 1: Computational Phylogenetic methods**

| Methods | Exhaustive search | Stepwise clustering |
|---|---|---|
| Character State | **M**aximum **p**arsimony **MP** | |
| | **M**aximum **l**ikelihood **ML** | |
| Distance Matrix | Fitch-Margoliash | UPGMA |
| | | **N**eighbour-**j**oining **NJ** |

## Objectives

The objectives of this paper are to design, implement and evaluate a new model of measuring the TCP performance of real time network based on bio-computing.
Furthermore the research aims to run the proposed model efficiently by exploiting:
 i. The windows and **J**ava **D**atab**a**se **C**onnectivity **JDBC** facilities (manipulate / control of this ultimate size of received and processed TCP packet, Sockets, protocols).
 ii. The computer simulation of bio-computing to obtain high TCP performance via real time network represented by phylogenetic tree (minimum tree) using phylogenetic analysis and its distance matrix strategy (NJ).
 iii. Appling proposed algorithm of Little's law (steady state) for single server queue to obtain TCP performance and comparing with bio-computing results.

# Client Server Architecture

The basic idea behind the client-server model of network communication is shown in figure (3). The client sends messages to the server requesting service of any kind. The server responds with messages containing the desired information or takes other appropriate action. The message containing the client request is encapsulated inside a network packet and transmitted over a physical connection to the server. Conceptually, a logical connection also exists between the client and server.

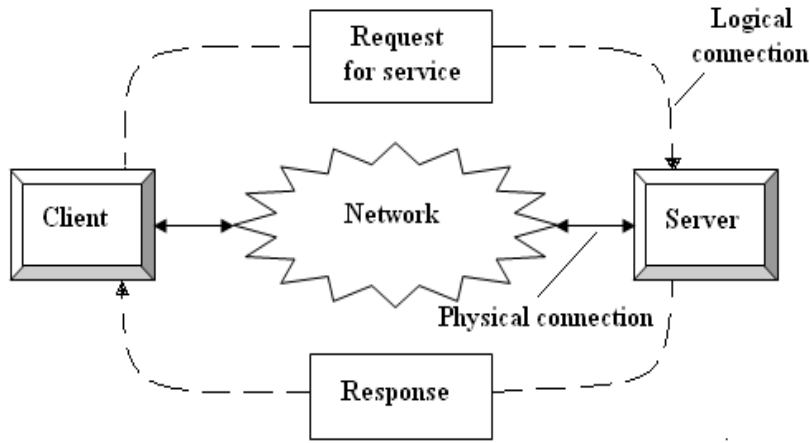

**Figure 3: Client-Server network architecture**

The client-server messages can easily be text based and look just like the examples. Or they can contain data. The server must already be running for the client to communicate with it. In addition, the client must know the IP address or domain name of the server to initial communication **[3]**.

There are two types of client server architecture, a simple server client and multiple server clients.

# Phylogenetic Tree

A phylogenetic tree, also called an evolutionary tree is a graphical representation of the evolutionary relationship between taxonomic groups. The term phylogeny refers to the evolution or historical development of a plant or animal species, or even a human tribe or similar group. Taxonomy is the system of classifying plants and animals by grouping them into categories according to their similarities **[4]**.

External (terminal) nodes are called OTUs and internal nodes are called Hypothetical Taxonomic Unit HTUs. A group of taxonomy is called a cluster; as shown in figure (4; A) the taxonomy A, B, C forms a cluster, having a common ancestor. The branching pattern that is the order of the nodes is called topology of the tree.

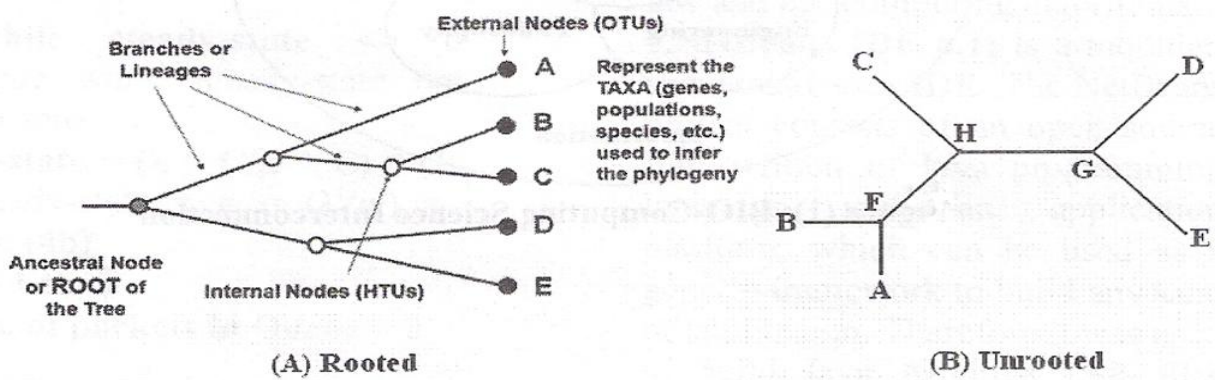

**Figure 4: Structure of (A) Rooted / (B) Unrooted phylogenetic tree**

A rooted phylogenetic tree is a directed tree with a unique node corresponding to the most recent common ancestor of all the entities at the leaves of the tree. The most common method for rooting trees is the uses of an uncontroversial outgroup.

Unrooted trees illustrate the relatedness of the leaf nodes without making assumptions about common ancestry. While unrooted trees can always be generated from rooted one by simply omitting the root, a root cannot be inferred from an unrooted tree without some means of identifying ancestry; this is normally done by including an outgroup in the input data or introducing additional assumptions about the relative rates of evolution on each branch. In Figure 2-3B A to E is called leaves. F to H is inferred nodes corresponding to ancestral species or molecules. Branches are also called edges **[5]**.

Figure (5) clarifying the nature of relationships between bio-computing at molecular- level technology and network routing model. The proposed model define and employ the mentioned relationships to produce an advance hybridized model, the proposed model is based on bio-computing mathematical level, which works at molecular level, represented in phylogenetic-NJ method equations.

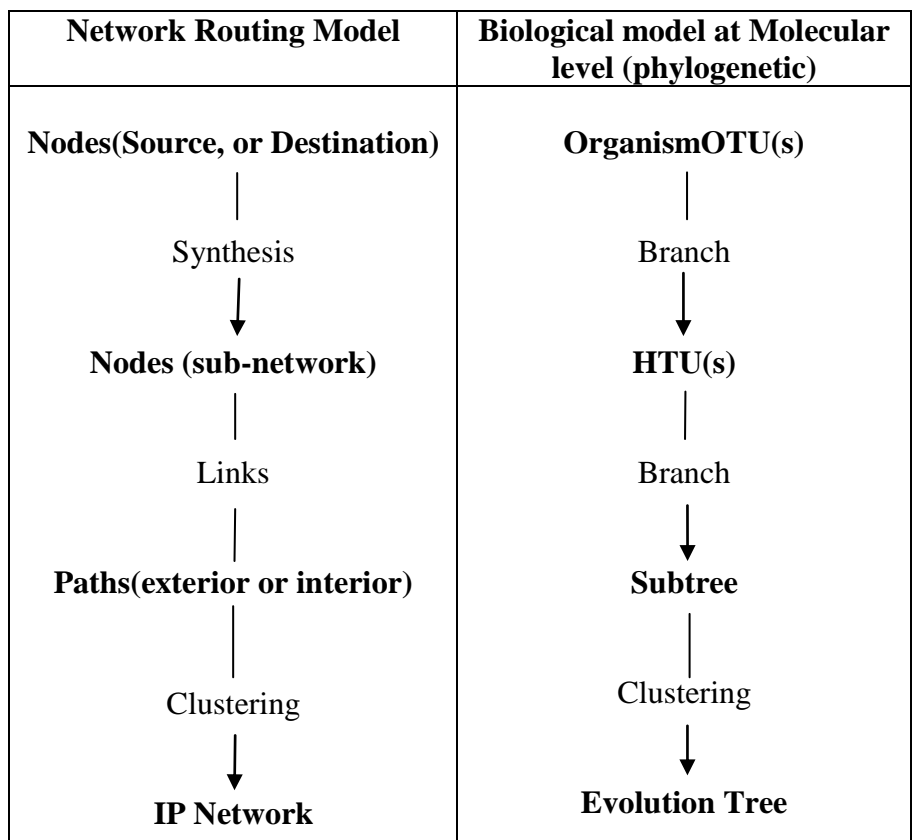

**Figure 5: Nature relationship among explained models [1]**.

## Proposed Technique for TCP Performance Measuring

To design, implement and evaluate a new model of measuring TCP performance, two protocols were selected; IRCD and FTP. IRCD is one of the most popular and interactive services of the internet that can exchange text messages interactively with other clients and FTP is a protocol designed to enable reliable transfer of files between stations.

Since the message and file exchange between users requires a centralized supervision, it needs a server for controlling the transition of messages and maintaining most of resources for the services then joining the server with database using JDBC for storing, retrieving, removing, and updating the data. The data mainly includes (header, user information, invited users, messages, bandwidth, number of hops, and others). Therefore it needs two applications client side and server side; these two applications are communicate over socket (Socket is one end-point of a two way communication link between two programs running on the network. Socket classes are used to represent the connection between a client program and a server program. The java.net package provides two classes: Socket and ServerSocket that implement the client side of the connection and the server side of the connection, respectively) by receiving and transmitting messages which carries specified meanings and works as a protocol between them.

The server side program listens to the client side requests and tries to serve them depending on received messages. The design of the server side program is differ than the client side program, the client side should be user friendly and the server side should be efficient in managing resources more than being user friendly.

A real network as a client and server was implemented which is node connected by node, it is possible to consider each path to be a tree therefore this tree is known as phylogentic tree. Neighbor joining method was used for reconstructing phylogentic tree, and then this can be taken as filter matrix which is relationship between links and paths. This filter matrix is able to select the links that are used or not. Filter matrix used for robustness which is routing algorithm property, robustness (Adaptation) with respect to failures and changing conditions (failure, congestion), then there are mathematical equations which depend on aggregate response and capacity to measure system performance. System performance was also measured by applying little's law which depends on arrival rate and service rate as shown in figure (6).

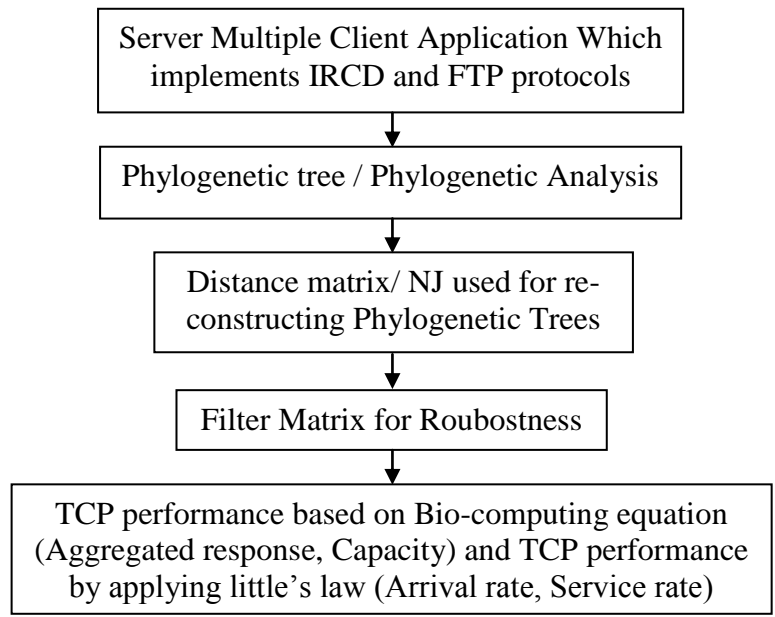

**Figure (6): Steps of TCP Performance Measuring.**

## Biological Kurd Messenger BIOKM Server and Client Application

For TCP performance measuring will need some application techniques the process is as shown below:
1. **Kurd Messenger Database KMDB Connectivity:** It has been estimated that half of all software development involves client/server operations and most client-server operations involve database connectivity. A database system is a repository or store of data. Database systems organize data in an orderly way making the data easily accessible and updatable. Instead of using a database, an ordinary text file can be used to store data, but a database management system makes updating, organizing and accessing data more efficient and user-friendly.

   Java supports database programming. JDBC is designed to provide Java programmers with a uniform way of accessing database systems. With JDBC one can access almost all current database systems such as Microsoft Structured Query Language SQL Server, Microsoft Access.

JDBC works like this: database vendors provide drivers for their particular database system to work with JDBC driver manager. JDBC provides an abstraction layer on top of these drivers. Programs written according to JDBC API would talk to the JDBC driver manager, which in turn, would use the drivers that it has to talk to the actual database [6].

2. **Kurd Messenger Internet Relay Chat Daemon KMIRCD:** There are two types of client server architecture, a simple server client and multiple server clients' architecture. Multiple server clients' architecture is used in this paper by using multithreading.
3. **Kurd Messenger FTP KMFTP:** The ability to transfer a file between computers is provided by the FTP. FTP can transmit or receive text or binary files. The primary function of FTP is defined as transferring files efficiently and reliably among host computers. FTP is a client-server application that uses two ports on both the client and the server. One port is used to exchange FTP control or command information and the other port is used to transfer the data **[3]**.
4. **Kurd Messenger Phylogenetic Tree:** Phylogenetic tree is a graphical representation of the evolutionary relationship between server and clients, starts to check database and take all information to draw the phylogenetic tree from database which are connected with the sever.
5. **Kurd Messenger Monitoring:** To monitor Kurd messenger, TCP/IP packet sniffer used which allows capturing TCP/IP packets that pass through the network adapter, and view the captured data as sequence of conversations between clients and server by using Jpcap which is a Java library for capturing and sending network packets, and WinPcap which is capture TCP/IP packets on all Windows operating systems.

## Normalization

Normalize the hybridized model (mathematical + biological) based on NJ method to become more adapted to simulate a computer network. The algorithm is as shown bellow **[2]**:

---

**BIOKM Hybridized model (Mathematical + Biological) Algorithm**
**Input a** {a: Source node}
**Input b** {b: Destination node}
**Input D** {D: path between two network nodes each element denoted by d(a,b), row of filter matrix}
**Input L** {L: Actual links (open/close) between two nodes, each element denoted by 1, column of filter matrix}
**R** {Filter or normalize matrix which is relationship between D and L defined as D Uses L if and only if R (d(a,b), l) = 1}
$R^T$ {Inverse relationship for the ordered pairs resulting from the filter matrix R only exchange column with rows}
**Range** {The range for the solution which is the first position of the ordered pairs (D,L)}
**Domain** {The domain for the solution which is the second position of the ordered pairs (D,L)}
**OPL** {Output packet length which is determined in TCP/IP monitor}
$T_T$ {Total time which is current time − server start time}
$T_S$ {Service time which is client departure time − client start time}
**BS** {Byte size that arrive per time unit which is equal to OPL / $T_T$}
**C** {Total capacity of links which is equal to (OPL / $\sum_{C=1}^{N} TS$)}
$D_{ab}$ {Distance between OTUs a and b}
$$Ui = \sum_{\substack{j=1 \\ j \neq i}}^{N} Dij \quad \{(Equation\ (1)\}$$
**Total aggregate response Q = Ui(Xi) subject to BS <= $C$** {Equation (2)}
**Print R, $R^T$, BS, C, Range, Domain, and Q**

---

## Little's Law of a Single Server Queue

A fundamental and simple relation with broad application is Little's Law; it can be applied to almost any system that is statistically in steady state. Little's law used in this research work as a comparison with bio-computing technique. The general and simplest setup is that the items arrive at an average rate of (λ) items per unit time. The items stay in the system an average of (W) units of time; finally, there is an average of (L) units in the system at any one time. Steady state system is Little's Law which relates these three variables L, λ, W as L = λ W.

Use a powerful known result, as little's queue formula, or recall that the M/M/1 queues in system. It has exponential interarrival times and single server with exponential service time. First will define the following quantities **[7]**:

- $\lambda$ = Average number of arrivals entering the system per unit time.
- **L** = Average number of packets present in the system.
- $L_q$ = Average number of packets waiting in queue.
- $L_s$ = Average number of packets in service.
- **W** = Average time a message spends in the system.
- $W_q$ = Average time a packet spends in queue.
- $W_s$ = Average time a packet spends in service.

Define: $\rho = \lambda / \mu$                              Equation (3)

Where ρ is called the traffic intensity (busy) of queuing system and $\pi_0 = 1-\rho$ called %idle. And steady state probability that j packets will be present in $\pi_j$ computed as follows:

$\pi_j = \rho^j (1-\rho)$                     Equation (4a)

Since $\pi_0 = 1-\rho$ and $\rho = \lambda / \mu$ so the result will be:

$\pi_j = \rho^j \pi_0$ or $\pi_j = (\lambda^j / \mu^j) \pi_0$       Equation (4b)

The average number of packet present in the queuing system (L) is given by:

**L** = ρ / 1-ρ or **L** = λ / μ-λ            Equation (5)

The expected number of packets waiting in queue is denoted as $L_q$ and compute as follows:

$L_q = \rho^2 / 1-\rho$ or $L_q = \lambda^2 / \mu (\mu-\lambda)$     Equation (6)

The expected number of packets in service is represented as $L_s$ obtained from:

$L_s = \rho$ or $L_s = \lambda / \mu$                  Equation (7)

And W, $W_q$, $W_s$ respectively as follows:

**W** = L / λ or **W** = 1 / μ-λ              Equation (8)

$W_q = L_q / \lambda$ or $W_q = \lambda / \mu (\mu-\lambda)$     Equation (9)

$W_s = L_s / \lambda$                          Equation (10)

The algorithm of Little's Law for a single server queue is shown below:

> **Algorithm of Little's Law (steady state) for Single server Queue [10]**
>
> **Input P** {P: Number of out packet in the system which is determined in TCP/IP monitor}
> **Input $T_T$** {$T_T$: Total Time which is current time – server start time}
> **Input μ** {μ: average number of service completion rate for packets per unit time}
> $\lambda = P / T_T$          {λ: average number of packets (arrival rate) per unit time}
> $\rho = \lambda / \mu$          {ρ: Utilization/Traffic intensity, equation 3}
> $L = \rho / (1-\rho)$          {equation 5}
> $L_q = \rho^2 / (1-\rho)$          {equation 6}
> $L_s = \rho$          {equation 7}
> $W = L / \lambda$          {equation 8}
> $W_q = L_q / \lambda$          {equation 9}
> $W_s = L_s / \lambda$          {equation 10}
> **%idles = (1-ρ)**          {equation 3}
> **Print L, $L_q$, $L_s$, W, $W_q$, $W_s$, and idles** {quantities of Little's law (steady state)}
> **C = 0**          {C: counter for the number of packets}
> **Co while steady-state <> 0**      {continue while Steady-state not equal to zero}
> **Steady-state = $(\lambda \wedge C / \mu \wedge C) * (1-\rho)$** {Steady-state = $\pi_j = (\lambda^j/\mu^j) \pi_0$, equation (4b) above}
> **If C <= 1 then**
>     **No. of packets in Queue = 0**
> **Else**
>     **No. of packets in Queue = C-1**
> **Loop** {return to $C_0$ while statement above}

## The tools and languages which are used

The tools which have been used in this research can be divided into:

A. **Software tools:** The software tools include:
   i. **Java Development kit JDK6:** is defined as a simple, object-oriented, distributed, interpreted, robust, secure, architecture neutral, portable, high-performance, multithreaded, and dynamic language. Therefore Java used for all research programs including client and server side programming. **[8, 9, 10]**
   ii. **TextPad:** is a popular text editor for the Microsoft Windows family of operating systems, used as a text editor.
   iii. **Microsoft Access:** is the top-notch database management system for all information management needs, from a simple address list to a complex inventory management system. It offers all the necessary tools for storing, retrieving, and interpreting the data. Therefore it was used for previous purposes **[11]**.
   iv. **Microsoft Excel:** is a proprietary spreadsheet application written and distributed by Microsoft. It features calculation, graphing tools, and table. Used as spreadsheet to achieve the computations of Little's law and bio-computing algorithms
   v. **NetBeans IDE 6.1:** is a modular, standards-based IDE. The NetBeans project consists of an open source IDE written in Java programming language and an application platform, which can be used as a generic framework to build any kind of application. Therefore it was used to build Java ARchive JAR files used for aggregating many files into one; it is based on the ZIP file format.
   vi. **Install Creator:** is a professional tool to create software installations. It offers a wizard interface that let's selects the files to include in the package, specify installation paths and then compiles the complete installation package into a compressed EXE. Therefore it was used to create KMSS and KMCS application installation and uninstall.

vii. **Quick Screen Capture:** is an all-in-one tool for screen capturing, image editing and organization. It can capture any part of the screen precisely in flexible ways, Used to capture the screen.

B. **Hardware tools:** The hardware tools include Networking (Server & Clients) components. Server-client is a type of network topologies that called star-topology.
   i. **Server Computer:**
      Usually called mainframe or mother computer that have a good performance and stability it used in the networking for serving other client(s). Microsoft Windows Server family must be installed on the server computer. Its installation is also like installing Microsoft windows XP Professional on the PCs or client computers.
   ii. **Client Computer :**
      Is a workstation computer or is said to be PCs that add to a domain or sends the request to a server and gets the server responses. Client computer may be any type of computers that runs an operating system like Microsoft Windows XP professional.
   iii. **Switch (Hub):-**
      Switch (Hub) is an electronic device that connects multiple computers to a server via RJ45 thought **LAN** cables. Figure (7) shows the relationship between server and client computers.

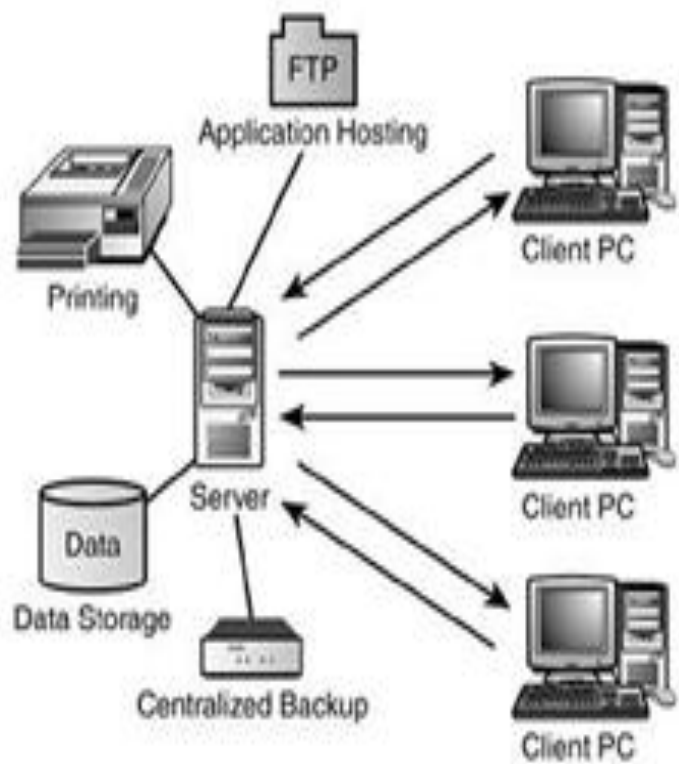

**Figure 7: The relationship between client and server computers**

## Results and Discussions

After installing BIOKM application which is consists of two main applications KMSS and KMCS then implementing it the performance value of IRCD application, FTP application, and IRCD & FTP applications are shown in table (2). The average performance values of IRCD, FTP, and IRCD&FTP were determined and used later in bio-computing and little's law techniques to determine TCP performance.

**Table 2: Performance Values using IRCD, FTP, IRCD and FTP Application**

| Measurement Factors | | IRCD | FTP | IRCD & FTP |
|---|---|---|---|---|
| No. of Online Clients | | 2 | 2 | 2 |
| No. of Servers | | 1 | 1 | 1 |
| Packet Sent (Packet) | | 6874 | 4612 | 7341 |
| Packet Sent Length (Byte) | | 6426 | 4304 | 6865 |
| Packet Received (Packet) | | 6452 | 4067 | 6832 |
| Packet Receive Length (Byte) | | 5868 | 3653 | 6214 |
| Total Arrival Time (Mili Second) | | 32484 | 32656 | 35438 |
| Total Departure Time (Mili Second) | | $1.22*10^{12}$ | $1.22*10^{12}$ | $1.22*10^{12}$ |
| Total Service Time (Mili Second) | | 73328 | 111922 | 113625 |
| Total Time (Mili Second) | | 111687 | 152297 | 155672 |
| Arrival Rate (Packet/Second) | | 57.8 | 26.7 | 43.9 |
| Service Rate (Packet/Second) | | 87.9 | 36.3 | 60.1 |
| Byte Size (Bit/Second) | | 420.3 | 191.9 | 319.3 |
| Capacity (Bit/Second) | | 640.2 | 261.1 | 437.5 |
| Bio-Computing | Total Aggregate Response | 0.6565 | 0.7349 | 0.7298 |
| | Expected Idle Time | 0.3435 | 0.2651 | 0.2702 |
| Little's Law | Traffic Intensity/Utilization | 0.6575 | 0.7355 | 0.7304 |
| | Expected Idle Time | 0.3425 | 0.2645 | 0.2996 |

## 1. Implementing Hybridized Model Algorithm

To implement hybridized model which is biological and mathematical model based on NJ method, consider the tree shown in figure (8).

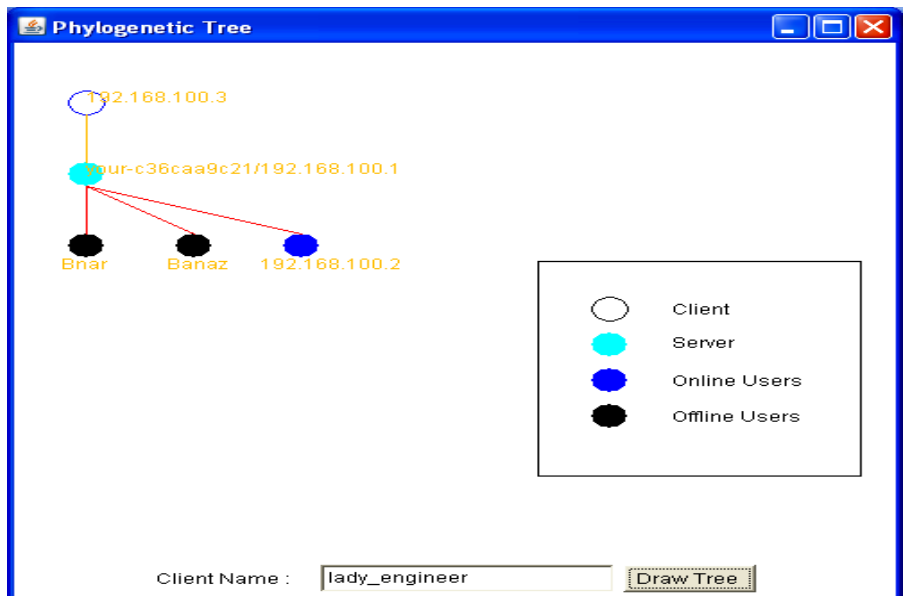

**Figure 8: Phylogenetic Tree for user lady_engineer**

Table (3) and table (4) show Filter or normalize matrix (R) and transport R ($R^T$) for the tree shown in figure (8) used for robustness:

$R_{li} = \begin{cases} 1 & \text{if link l uses source i} \\ 0 & \text{otherwise} \end{cases}$

**Table 3: Matrix R for User lady_engineer**

| Matrix R | | |
|---|---|---|
| | Paths | |
| Links | R | P 1 |
| | L 1 | 1 |
| | L 2 | 0 |
| | L 3 | 0 |
| | L 4 | 1 |

**Table 4: Transport Matrix R ($R^T$) for User lady_engineer**

| Inverse Matrix R ($R^T$) | | | | | |
|---|---|---|---|---|---|
| | | Links | | | |
| Paths | $R^T$ | L1 | L2 | L3 | L4 |
| | P1 | 1 | 0 | 0 | 1 |

Depending on equation 2 shown in BIOKM hybridized model algorithm and results in table (2) the results of biological performance based on NJ method are shown in table (5) by using Microsoft Excel as an efficient spreadsheet tool.

**Table (5): Biological Performance Based on NJ Method**

| Biological Performance (Throughput) Based on NJ Method | |
|---|---|
| Average Byte size arrive per time unit BS | 310.5 bit/second |
| Capacity C per time unit | 446.3 bit/second |
| Total aggregate response | 0.695720367 |
| Expected % idle time = 1 – Total aggregate response | 0.304279633 |
| # of Servers | 1 |

## 2. Implementing little's Law Algorithm of Single Server Queue

Applying the proposed algorithm of Little's law (steady state) for single server queue, by using Microsoft Excel as an efficient spreadsheet tool which can be used to achieve the computations of this algorithm (obtained results of performance), as shown in figure (9).

**Figure 9: Steady-State Performance**

| | A | B | C | D | E |
|---|---|---|---|---|---|
| 1 | Steady-State Performance (Throughput) of single Server Queuing (Little's Law L=λ.W) | | Messages | Steady-State Probability | Expected Idle Time |
| 2 | | | | $(\lambda^C /\mu^C)(1-P)$ | |
| 3 | Average no. of Packets Arrived Per second (arrival rate) λ | 42.8 | 0 | 0.302931596 | 0.302931596 |
| 4 | Average No. of service Completion for Packets per Second (Service rate) μ | 61.4 | 1 | 0.211164044 | 0.302931596 |
| 5 | Traffic Intensity/Utilization (Busy) P = λ / μ | 0.697068404 | 2 | 0.147195783 | 0.302931596 |
| 6 | Average No. of Packets in System = L = p / 1 – p | 2.301075269 | 3 | 0.10260553 | 0.302931596 |
| 7 | Average No. of Packets Waiting in Queue = Lq = p2 /1 – p | 1.604006865 | 4 | 0.071523073 | 0.302931596 |
| 8 | Average No. of Packets in Service = Ls = p | 0.697068404 | 5 | 0.049856474 | 0.302931596 |
| 9 | Average Time a Packets Spends in System = W = L / λ | 0.053763441 | 6 | 0.034753373 | 0.302931596 |
| 10 | Average Time a Packets Spends in Queue = Wq = Lq / λ | 0.037476796 | 7 | 0.024225478 | 0.302931596 |
| 11 | Average Time a Packets Spends in Service = Ws = Ls / λ | 0.016286645 | 8 | 0.016886815 | 0.302931596 |
| 12 | Expected % idle time = 1 – p | 0.302931596 | 9 | 0.011771265 | 0.302931596 |
| 13 | # of Servers | 1 | 10 | 0.008205377 | 0.302931596 |

The computation of Little's law (performance and throughput) appears in spreadsheet shown in figure (10), while the charts are shown in figure (11).

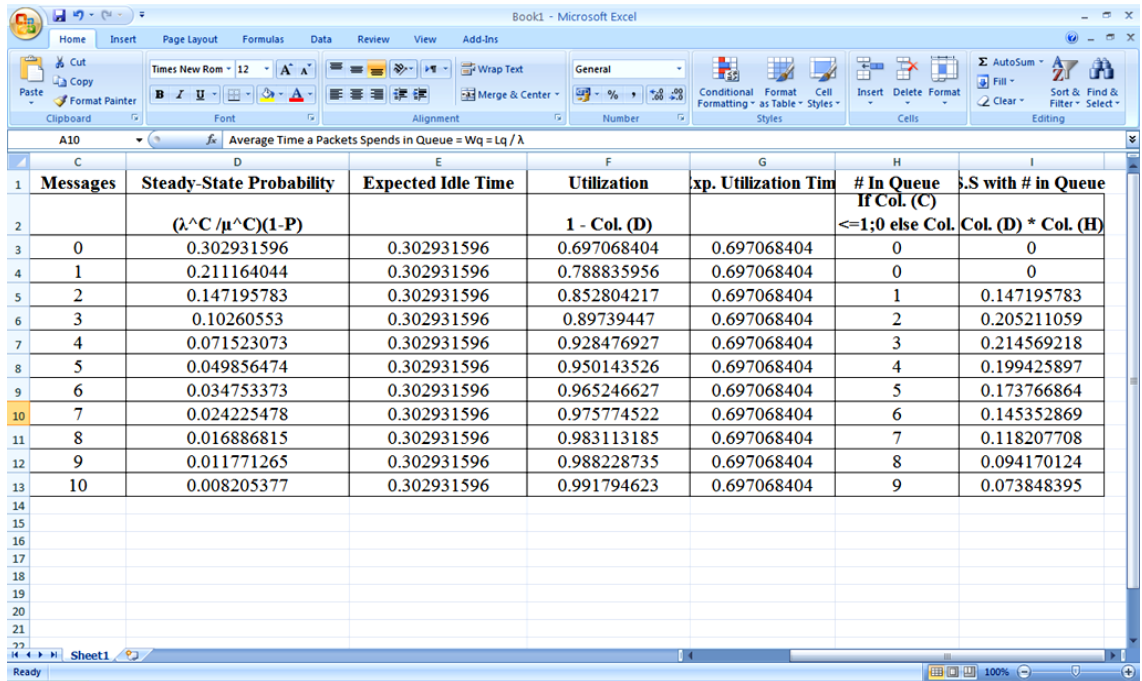

**Figure 10: Reveal computation of steady state (performance and throughput)**

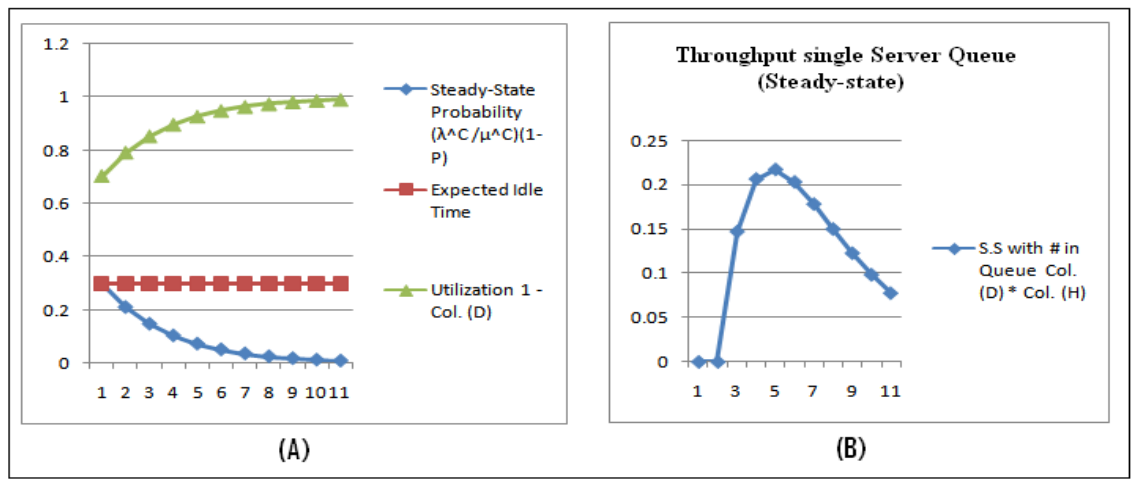

**Figure 11: Explain charts of steady state of single server queue**

Figure (11; A) represents the probability of percentage utilization of Little's law (steady state) and opposite percentage idles of single server queue. The server percentage utilization reach nearest to percentage 100 when the number of messages in queue is equal or greater than 10, while percentage idle becomes zero. Figure (11; B) shows the throughput of single server queue.

### Conclusions

The conclusions obtained from this research work can be stated as follows:
i. BIOKM phylogenetic tree implementation showed that NJ method produces a unique find tree under the principle of minimum evolution, and its efficiency in obtaining the correct tree.
ii. Jpcap is a good java library for capturing and sending network packets, and Winpcap is a good capture TCP/IP packet on all windows operating system.

iii. Java programming language is more applicable and organized language than any other languages in this field and Java programmable model can be used to solve different practical real problems.
iv. Measuring TCP performance using bio-computing technique especially molecular calculation provides wisdom results and it is possible to exploit all facilities of phylogenetic analysis. It shows very good performance and throughput as shown in table (5).
v. The Little s Law (Steady State) technique which is implemented for single server queue shows very good performance and throughput as shown in figures (9), (10) and (11, when the number of messages increases, the steady-state probability decreases, utilization increases, and steady-state probability with number in queue decreases. All these mean that the performance will increase.
vi. The results obtained from bio-computing and little's law techniques are shown in table (6).

Table 6: Comparison between results of bio-computing and little's law

| Technique / Average Performance | Bio-computing | Little's Law | Difference Percentage |
|---|---|---|---|
| Utilization | 0.695720367 | 0.697068404 | 0.193386616% |
| Expected idle time | 0.304279633 | 0.302931596 | 0.443025708% |

Table (6) shows the TCP performance measuring based on bio-computing and little's law which is a famous performance measuring technique, the results are very close to each other when the average values were taken into consideration, the utilization (Busy) or traffic industry obtained from bio-computing was 69.57%, while utilization obtained from little's law was 69.70%, the difference percentage was 0.193386616 this is because of local implementation. Best results of bio-computing are expected in the case of remote server; hence bio-computing is a very efficient technique of measuring TCP performance.

vii. There are many factors affecting BIOKM such:
- Dynamic parameter of network such as data rate and speed.
- Constant factors such as antivirus and memory size which make the system to be slower.
- Hub structure such as hub only or hub plus firewall. Firewall doesn't allow KMCS to connect with KMSS.
- Others such as Central Processor Unit (CPU) temperature which make the system to be slower.

## Suggestions

The suggestions for further studies can be stated as follow:
i. Measuring TCP performance in the case of remote server after the availability of required concepts such as server account and authorization to know the behavior of the server.
ii. Using DNA level applying mutation types for error correction, in the network Cyclic Redundancy Code (CRC) and Huffman code used for error detection and correction.
iii. Updating BIOKM by applying additional components such as offline send massage, chat room, voice chat, and web camera by studying addition subjects in java programming language.